\documentclass[11pt]{article}
\usepackage{amsmath,amssymb,epsfig,bm}

\date{\empty}

\begin{document}

\title{\bf The deceleration parameter in `tilted'\\ Friedmann universes}

\author{Christos G. Tsagas and Miltiadis I. Kadiltzoglou\\ {\small Section of Astrophysics, Astronomy and Mechanics, Department of Physics}\\ {\small Aristotle University of Thessaloniki, Thessaloniki 54124, Greece}}

\maketitle

\begin{abstract}
Large-scale peculiar motions are believed to reflect the local inhomogeneity and anisotropy of the universe, triggered by the ongoing process of structure formation. As a result, realistic observers do not follow the smooth Hubble flow but have a peculiar, `tilt', velocity relative to it. Our Local Group of galaxies, in particular, moves with respect to the universal expansion at a speed of roughly 600~km/sec. Relative motion effects are known to interfere with the observations and their interpretation. The strong dipolar anisotropy seen in the Cosmic Microwave Background, for example, is not treated as a sign of real universal anisotropy, but as a mere artifact of our peculiar motion relative to the Hubble flow. With these in mind, we look into the implications of large-scale bulk motions for the kinematics of their associated observers, by adopting a `tilted' Friedmann model. Our aim is to examine whether the deceleration parameter measured in the rest-frame of the bulk flow can differ from that of the actual universe due to relative-motion effects alone. We find that there is a difference, which depends on the speed as well as the scale of the bulk motion. The faster and the smaller the drifting domain, the larger the difference. In principle, this allows relatively slow peculiar velocities to have a disproportionately strong effect on the value of the deceleration parameter measured by observers within bulk flows of, say, few hundred megaparsecs. In fact, under certain circumstances, it is even possible to change the sign of the deceleration parameter. It goes without saying that all these effects vanish identically in the Hubble frame, which makes then an illusion and mere artifact of the observers' relative motion.
\end{abstract}

\section{Introduction}
On large enough scales, of the order of several hundred megaparsecs and beyond, our universe appears uniform to a very good approximation. On sufficiently small scales, however, there is a lot of structure (in the form of galaxies, galaxy clusters, superclusters and voids), which destroys local homogeneity and isotropy to a larger or lesser degree. Bulk peculiar motions are a direct result of the ongoing process of structure formation. Typical galaxies do not follow the smooth Hubble expansion but have a `drift' motion relative to it~\cite{TD}. Our Local Group of galaxies, for example, has a peculiar velocity of approximately 600~km/sec with respect to the frame of the Cosmic Microwave Background (CMB). The latter is defined as the coordinate system where the CMB dipole vanishes and provides the reference frame of the universal expansion, relative to which we can define and measure large-scale bulk motions.

Peculiar velocities are expected to fade away as we move on to progressively larger wavelengths. The current concordance cosmological model, namely the $\Lambda$CDM paradigm, allows for peculiar velocities up to 250~km/sec on scales close to 100/h~Mpc, with decreasing magnitude as we move on to larger lengths.\footnote{The dimensionless variable $h=H/100$~km/sec\,Mpc defines the normalised Hubble parameter} Taking cosmic variance into account could bring the aforementioned limit higher, though not to the magnitudes reported by a number of recent surveys. The work of~\cite{WFH}, in particular, claims bulk velocities about 400~km/sec within a sphere of $100/h$~Mpc size. Using different methods, however, other authors have reported lower amplitudes for the local bulk flow~\cite{LTMC,CMSS}, some of which lie within the general $\Lambda$CDM predictions~\cite{ND}. Much faster peculiar motions, between 1000~km/sec and 4000~km/sec, covering scales of Gpc-order have also been reported in the literature~\cite{KA-BKE}. Nevertheless, such very large-scale `dark flows', as they are commonly known, appear to be at odds with the Planck constraints. The latter set an upper limit of roughly 400~km/sec within regions larger than 500~Mpc~\cite{Aetal}, but the issue may not be over yet~\cite{A-B}.

Relative motions are known to interfere with the observations and with their interpretation. The strong dipolar anisotropy seen in the CMB spectrum, for instance, is not treated as a sign of real cosmic anisotropy, but as a mere artifact (a Doppler-like effect) of our motion with respect to the smooth Hubble expansion. With these in mind, we will employ a tilted Friedmann-Robertson-Walker (FRW) cosmology and a two-fluid approach to investigate the implications of large-scale bulk motions for the mean kinematics of the `drifting' observers. We introduce two timelike 4-velocity fields, tangent to the worldlines of their corresponding observers, which move relative to each other. The first 4-velocity field will be identified with the smooth Hubble flow and will define our frame of reference. Relative to this coordinate system, we will assume that the universe is an FRW spacetime filled with conventional pressureless matter. The second 4-velocity field, which has a small but finite peculiar velocity with respect to the first, will define the rest-frame of a realistic observer in a typical galaxy that moves relative to the Hubble expansion. Treating the peculiar velocity as a small fraction of the Hubble speed, on the corresponding scale, confines the domain of our study to wavelengths larger than $\sim100$~Mpc. The latter is also thought to mark the `homogeneity threshold' of our cosmos, namely the scale beyond which the universe behaves as an almost-FRW model, although this is a working hypothesis rather than a hard observational fact.

Assuming that the drifting observers reside within a large-scale bulk flow, like those reported in~\cite{WFH}-\cite{KA-BKE}, we focus on their average volume expansion and more specifically on the deceleration parameter measured in their own rest-frame. Following~\cite{T1,T2}, we find that bulk motions can modify the local expansion (Hubble) rate and that the magnitude of the change depends on the speed of the peculiar flow. Not surprisingly, and in agreement with~\cite{T1,T2}, we also find that variations in the local volume expansion change the associated deceleration/acceleration rate and therefore the local value of the deceleration parameter. To estimate these changes, we extend the work of~\cite{T1,T2} and turn to linear cosmological perturbation theory, which allows us to connect peculiar velocities to density perturbations. The result is a simple relation expressing the deceleration parameter, as measured in the rest-frame of an observer living inside a large-scale bulk flow, in terms of the deceleration parameter of the actual universe and gradients of the peculiar velocity field.

Perhaps the most interesting feature of the aforementioned mathematical expression is its scale-dependence. We find that the effect of relative motion on the deceleration parameter, as measured in the rest-frame of the bulk flow, increases with decreasing scale. The threshold is the Hubble radius, which makes coherent drift motions with sizes of few hundred Mpc particularly `vulnerable' to relative-motion effects. On these scales, even the peculiar velocities allowed by the $\Lambda$CDM model can seriously affect the local value of the deceleration parameter. In addition to the magnitude of the velocity and the size of the bulk motion, the impact on the deceleration parameter also depends on whether the peculiar flow is (slightly) contracting or expanding. Crucially, when the volume of the bulk is decreasing, relative motion effects can even change the sign of the deceleration parameter there. None of these effects is `real', of course, since they all vanish in the CMB-frame. They are mere artifacts of the observers' peculiar motion, just like the CMB dipole, but ignoring them could potentially lead to a serious misinterpretation of the observational data. Put another way, measuring a negative deceleration parameter in the rest-frame of a bulk flow may not necessarily mean that the universe itself is accelerating.

\section{Aspects of relative motion}\label{sARM}
\begin{figure}[tbp]
\begin{center}
\includegraphics[height=2.5in,width=5.5in,angle=0]{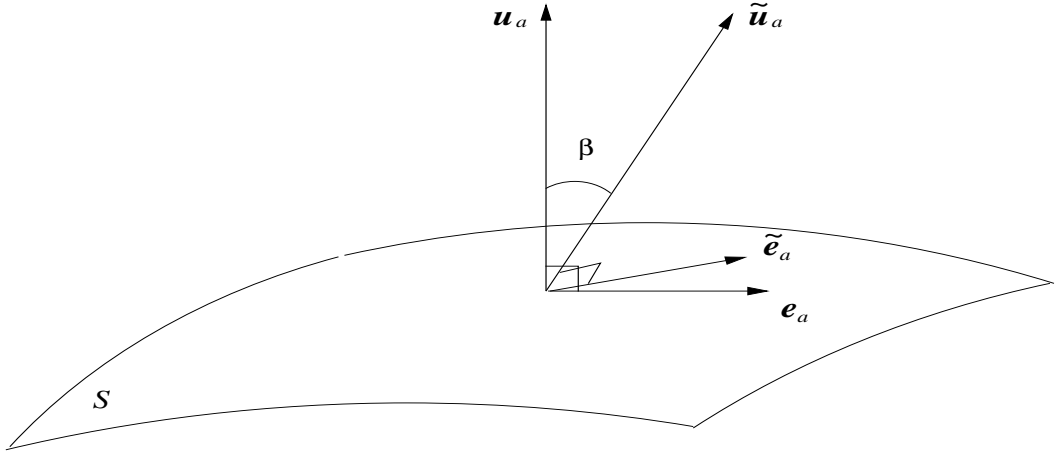}\quad
\end{center}
\caption{Observers with 4-velocity $\tilde{u}_a$ have peculiar velocity $v_a=ve_a$ relative to the $u_a$-field. By construction, $u_ae^a=0=\tilde{u}_a\tilde{e}^a$ and $e_ae^a=1= \tilde{e}_a\tilde{e}^a$. The hyperbolic angle $\beta$, with $\cosh\beta=-u_a\tilde{u}^a$, defines the `tilt' between $u_a$ and $\tilde{u}_a$, while $S$ is the 3-D space normal to $u_a$~\cite{KE}.}  \label{mfluid1}
\end{figure}

Consider two families of relatively moving observers, with worldlines tangent to the (timelike) 4-velocity fields $u_a$ and $\tilde{u}_a$ respectively. Assuming that $v_a$ is the peculiar velocity of the latter family with respect to the former, we have
\begin{equation}
\tilde{u}_a= \gamma(u_a+v_a)\,,  \label{Lorentz1}
\end{equation}
where $\gamma=(1-v^2)^{-1/2}>1$ is the associated Lorentz factor. Note that $u_av^a=0$, $v^2=v_av^a$ and $u_au^a=-1= \tilde{u}_a\tilde{u}^a$ by construction (e.g.~see~\cite{KE}). The above relation also determines the `tilt' angle between the two 4-velocity fields. In particular, the inner product $u_a\tilde{u}^a=-\gamma<-1$, defines the hyperbolic angle ($\beta$) between $u_a$ and $\tilde{u}_a$ (see Fig.~\ref{mfluid1}), so that $\cosh\beta=-u_a\tilde{u}^a=\gamma>1$ and $\beta=\ln(\gamma+\sqrt{\gamma^2-1})$. In what follows we will only look at non-relativistic peculiar motions, with $v^2\ll1$ and $\gamma\simeq1$. Therefore, in our case, the Lorentz boost reduces to
\begin{equation}
\tilde{u}_a= u_a+ v_a\,,  \label{Lorentz2}
\end{equation}
while the tilt angle between $u_a$ and $\tilde{u}_a$ becomes very small (i.e.~$\beta\simeq\ln1\simeq 0$).

The velocity fields $u_a$ and $\tilde{u}_a$ define the time-direction of their associated observers, while their 3-dimensional rest-spaces are normal to the corresponding 4-velocities. Spatial geometry is determined via the so-called projection tensors. These are defined by $h_{ab}=g_{ab}+u_au_b$ and $\tilde{h}_{ab}=g_{ab}+\tilde{u}_a\tilde{u}_b$, where $g_{ab}$ is the metric of the whole spacetime, so that $h_{ab}u^b=0=\tilde{h}_{ab}\tilde{u}^b$.\footnote{Irrotational 4-velocity fields are hypersurface orthogonal as well. Then, the projector also acts as the metric tensor of the observers' 3-D space. In this study the $u_a$-field will be irrotational by construction (see \S~\ref{sL-SPMs} below).} In addition, $h_{ab}=h_{(ab)}$, $h_{ab}=h_{ac}h^c{}_b$ and $h_a{}^a=3$, with exactly analogous relations holding in the `tilted' frame as well (e.g.~see~\cite{E,TCM}). Finally, employing the aforementioned 4-velocity fields and their corresponding projectors, we can define temporal and spatial derivatives, relative to the two frames. Thus, from now on, overdots and primes will indicate time differentiation in the $u_a$-frame and its tilted counterpart respectively (i.e.~${}^{\cdot}=u^a\nabla_a$ and $~{}^{\prime}= \tilde{u}^a\nabla_a$, with $\nabla_a$ representing 4-dimensional covariant derivatives). The 3-D (spatial) covariant-derivative operators, on the other hand, will be  ${\rm D}_a=h_a{}^b\nabla_b$ (orthogonal to the $u_a$-field) and $\tilde{{\rm D}}_a= \tilde{h}_a{}^b\nabla_b$ (normal to $\tilde{u}_a$).

We should point out that the peculiar velocity field ($v_a$) lies in the 3-D space of the $u_a$-frame but not in that of tilted observers. In particular, one can easily show that $\tilde{u}_av^a=\gamma v^2\neq0$ and that $\tilde{h}_a{}^bv_b= v_a+\gamma v^2\tilde{u}_a\neq v_a$. Of course, when $v^2\ll1$ and $\gamma\simeq1$, the tilt angle ($\beta$) tends to zero and the aforementioned 3-D spaces essentially coincide (i.e.~$\tilde{e}_a\simeq e_a$ -- see Fig.~\ref{mfluid1}).

\section{Kinematics of peculiar motions}\label{sKPMs}
Once the spacetime has been split into time and 3-D space, one can chose a family of observers and proceed to decompose every variable, every operator and every equation into their temporal and spatial parts relative to the chosen frame. For instance, the covariant derivative of the $u_a$-field splits into the irreducible components of its motion according to (e.g.~see~\cite{E,TCM})
\begin{equation}
\nabla_bu_a= {1\over3}\,\Theta h_{ab}+ \sigma_{ab}+ \omega_{ab}- A_au_b\,.  \label{Nbua}
\end{equation}
Here, the (volume) scalar $\Theta=\nabla^au_a$ describes the mean separation between the observers, with positive values of $\Theta$ indicating expansion and negative ones contraction. The symmetric and trace-free tensor $\sigma_{ab}={\rm D}_{\langle b}u_{a\rangle}$ represents the shear and the antisymmetric tensor $\omega_{ab}={\rm D}_{[b}u_{a]}$ the vorticity of the observers' worldlines. Also, $A_a=\dot{u}_a= u^b\nabla_bu_a$ is the 4-acceleration vector and vanishes when these worldlines are timelike geodesics. Similarly, the covariant derivative of the $\tilde{u}_a$-field decomposes as
\begin{equation}
\nabla_b\tilde{u}_a= {1\over3}\,\tilde{\Theta}\tilde{h}_{ab}+ \tilde{\sigma}_{ab}+ \tilde{\omega}_{ab}- \tilde{A}_a\tilde{u}_b\,,  \label{tNbua}
\end{equation}
where now $\tilde{\Theta}=\nabla^a\tilde{u}_a$, $\tilde{\sigma}_{ab}= \tilde{{\rm D}}_{\langle b}\tilde{u}_{a\rangle}$, $\tilde{\omega}_{ab}=\tilde{{\rm D}}_{[b}\tilde{u}_{a]}$ and $\tilde{A}_a=\tilde{u}^{\prime}_a= \tilde{u}^b\nabla_b\tilde{u}_a$. Starting from these definitions, one can derive the expressions relating the tilded variables to their non-tilded counterparts, in terms of the gradients of the peculiar velocity field. Treating the peculiar velocities as perturbations on an FRW background, we arrive at the linear transformation laws~\cite{M,ET}
\begin{equation}
\tilde{\Theta}= \Theta+ \tilde{{\rm D}}^av_a= \Theta+ {\rm D}^av_a  \label{ltrans1}
\end{equation}
and
\begin{equation}
\tilde{A}_a= A_a+ v_a^{\prime}+ {1\over3}\,\Theta v_a= A_a+ \dot{v}_a+ {1\over3}\,\Theta v_a\,.  \label{ltrans2}
\end{equation}
Note that, in deriving the above, one needs to remember that $\tilde{h}_{ab}=h_{ab}+2u_{(a}v_{b)}$, to first order in $v$. This guarantees that $\tilde{{\rm D}}^av_a= {\rm D}^av_a$ at the same perturbative level. Also, using Eq.~(\ref{Lorentz2}), one can easily verify that $v_a^{\prime}=\dot{v}_a$ to linear order in $v$.

Relations (\ref{ltrans1}) and (\ref{ltrans2}) show that, even though the universe may be an exact FRW model relative to the $u_a$-frame, it will not appear so to the tilted observers. The former, in particular, ensures that relatively moving observers measure different expansion rates. The latter, on the other hand, shows that we cannot simultaneously treat the worldlines of the relatively moving observers as timelike geodesics, even when the peculiar velocities are small. More specifically, in general we have $\tilde{A}_a\neq0$ when $A_a=0$ (or vice-versa), due to relative motion effects alone.

\section{Large-scale peculiar motions}\label{sL-SPMs}
By construction, the expansion of the universe defines is a preferred reference frame, with respect to which one can define relative motions and measure peculiar velocities. This is the coordinate system of the smooth Hubble flow, namely the frame where the CMB dipole vanishes. In what follows we will always identify the $u_a$-field with the 4-velocity of an observer following the Hubble expansion. We will also assume that the aforementioned (fundamental) observer `sees' the universe as an FRW cosmology, containing conventional pressure-free dust. This will be the `background' universe of the tilted observers, namely those with 4-velocity $\tilde{u}_a$, living in a typical galaxy (like our Milky Way) that moves relative to the CMB frame with peculiar velocity $v_a$.

The rest of this paper will look into the average kinematics of the aforementioned two families of observers. More specifically, we will focus on the rate of their mean (volume) expansion and on their acceleration/deceleration. Since we are dealing with non-relativistic peculiar motions, the volume scalars measured in the two frames are related by~\cite{T1,T2}
\begin{equation}
\tilde{\Theta}= \Theta+ \tilde{\vartheta}\,,  \label{Thetas1}
\end{equation}
with $\tilde{\vartheta}=\tilde{\rm D}^av_a$ (see Eq.~(\ref{ltrans1}) above).\footnote{By construction, the two volume scalars may be written as $\Theta=3H$ and $\tilde{\Theta}=3\tilde{H}$, where $H$ and $\tilde{H}$ are the Hubble parameters measured in the $u_a$ and the $\tilde{u}_a$ frames respectively (e.g.~see~\cite{E,TCM})} Note that $\Theta>0$ always due to the background expansion (i.e.~that of the $u_a$-field). In addition, since we will be dealing with relatively slow peculiar flows, we will assume that $|\tilde{\vartheta}|\ll \Theta$. This makes $v_a$ and $\tilde{\vartheta}$ first-order perturbations. The above relation shows that $\tilde{\Theta}\neq\Theta$, that is the expansion rate of the tilted observer and that of the actual universe are generally different. In particular, when $\tilde{\vartheta}$ is positive, the tilted observers expand slightly faster than the actual universe, while in the opposite case (i.e.~for $\tilde{\vartheta}<0$) their expansion is slightly slower. One therefore expects to see a difference in the acceleration/deceleration rates between the two frames. Indeed, taking the time derivative of the above, relative to the $\tilde{u}_a$-frame, and keeping up to linear-order terms, we arrive at
\begin{equation}
\tilde{\Theta}^{\prime}= \dot{\Theta}+ \tilde{\vartheta}^{\prime}\,,  \label{dotThetas1}
\end{equation}
where $\tilde{\Theta}^{\prime}=\tilde{u}^a\nabla_a\tilde{\Theta}$, $\dot{\Theta}=u^a\nabla_a\Theta$ and $\tilde{\vartheta}^{\prime}= \tilde{u}^a\nabla_a\tilde{\vartheta}$~\cite{T1,T2}. Accordingly, $\tilde{\Theta}^{\prime}\neq\dot{\Theta}$, with their actual difference depending on the sign and on the magnitude of $\tilde{\vartheta}^{\prime}$. Note that, although current observations strongly suggest that $\tilde{\vartheta}\ll\Theta$ on scales greater than $\sim100$~Mpc, we should not a priori extend this condition to their time derivatives.

Different values for $\tilde{\Theta}^{\prime}$ and $\dot{\Theta}$ imply that the deceleration parameters, as measured in the two frames, will differ as well. By definition, we have
\begin{equation}
\tilde{q}= -\left(1+{3\tilde{\Theta}^{\prime}\over\tilde{\Theta}^2}\right) \hspace{15mm} {\rm and} \hspace{15mm} q= -\left(1+{3\dot{\Theta}\over\Theta^2}\right)\,, \label{qs1}
\end{equation}
for the deceleration parameters in the tilted and the CMB frame respectively. Solving the above for $\tilde{\Theta}^{\prime}$ and $\dot{\Theta}$ and substituting the resulting expressions into Eq.~(\ref{dotThetas1}), gives
\begin{equation}
1+ \tilde{q}= (1+q)\left({\Theta\over\tilde{\Theta}}\right)^2- {3\tilde{\vartheta}^{\prime}\over\tilde{\Theta}^2}\,.  \label{qs2}
\end{equation}
This provides our first relation between the deceleration parameter measured in the rest frame of a typical galaxy ($\tilde{q}$) and that of the actual universe ($q$). Moreover, recalling that $\tilde{\Theta}=\Theta+ \tilde{\vartheta}$ and performing a simple Taylor expansion reduces the above into
\begin{equation}
1+ \tilde{q}= (1+q)\left(1-2{\tilde{\vartheta}\over\Theta}\right)- {3\tilde{\vartheta}^{\prime}\over\Theta^2}\,,  \label{qs3}
\end{equation}
after keeping up to $v$-order terms. Therefore, in the absence of peculiar motions (i.e.~when $\tilde{\vartheta}=0= \tilde{\vartheta}^{\prime}$), the two deceleration parameters coincide (as expected). Generally, however, $\tilde{q}\neq q$ and their difference is mainly decided by $\tilde{\vartheta}^{\prime}$, namely by the time evolution of $\tilde{\vartheta}$. This scalar, which has been defined as $\tilde{\vartheta}=\tilde{\rm D}^av_a$, describes the mean expansion/contraction of the peculiar flow.\footnote{The irreducible kinematic variables of the bulk motion are obtained by decomposing the spatial gradient of the peculiar velocity field. In particular, relative to the tilted frame, we have
\begin{equation}
\tilde{\rm D}_bv_a= {1\over3}\,\tilde{\vartheta}\tilde{h}_{ab}+ \tilde{\varsigma}_{ab}+ \tilde{\varpi}_{ab}\,,  \label{tDbva}
\end{equation}
with $\tilde{\vartheta}=\tilde{\rm D}^av_a$, $\tilde{\varsigma}_{ab}=\tilde{\rm D}_{\langle b}v_{a\rangle}$ and $\tilde{\varpi}_{ab}=\tilde{\rm D}_{[b}v_{a]}$. Therefore, in analogy with $\Theta$ and $\tilde{\Theta}$, the scalar $\tilde{\vartheta}$ describes the average separation of the peculiar-flow lines, namely their expansion or contraction. Similarly, $\tilde{\varsigma}_{ab}$ represents the `peculiar' shear and $\tilde{\varpi}_{ab}$ is the `peculiar' vorticity relative to the tilted frame~\cite{ET,TK}.}

In principle, one could calculate the ratios $\tilde{\vartheta}/\Theta$ and $\tilde{\vartheta}^{\prime}/\Theta^2$ from the observations (recall that $\Theta=3H$ in the CMB frame) and then use expression (\ref{qs3}) to estimate the deceleration parameter in the rest-frame of the peculiar flow. In practice, however, the surveys only provide the mean velocity ($\langle v\rangle$) of the bulk. One could then use the (approximate) relation $|\tilde{\vartheta}|\simeq |\partial^av_a|\simeq3\langle v\rangle/\lambda$, where $\lambda$ are the dimensions of the bulk motion, to obtain an estimate for the absolute value of the associated volume scalar. The problem is that, so far at least, the data cannot provide the sign of $\tilde{\vartheta}$ on the scales of interest (i.e.~for $\lambda\gtrsim100$~Mpc). Even more difficult is to estimate the time derivative of $\tilde{\vartheta}$ (and therefore the ratio  $\tilde{\vartheta}^{\prime}/\Theta^2$) directly from the observations. Therefore, in the next section, we will attempt to `ascertain' $\tilde{\vartheta}^{\prime}$ theoretically.

\section{The Raychaudhuri equation of the bulk 
flow}\label{sREBF}
The Raychaudhuri equation monitors changes in the mean separation between the worldlines of neighbouring observers, that is whether these expand/contract at an accelerated or decelerated pace. Although Raychaudhuri's formula has been typically associated with timelike worldlines, it can be applied to curves of any type. For instance, changes in the average separation between the peculiar flow lines, which are spacelike curves, are monitored by what we may call the \textit{peculiar Raychaudhuri equation}~\cite{TK}. The latter can be obtained in the `traditional' way, namely by applying the Ricci identities to the peculiar velocity field, or by taking the time derivative of $\tilde{\vartheta}$ in the tilted frame. In either case, assuming non-relativistic peculiar flows in an FRW cosmology with dust, the calculation gives
\begin{equation}
\tilde{\vartheta}^{\prime}= -{1\over3}\,\Theta\tilde{\vartheta}+ \tilde{\rm D}^av_a^{\prime}\,,  \label{pRay1}
\end{equation}
to first order~\cite{TK}. Note that in deriving the above we have used two sets of linear relations between the variables in two frames~\cite{M}
\begin{equation}
\tilde{\rho}= \rho\,, \hspace{10mm} \tilde{p}= 0\,, \hspace{10mm} \tilde{q}_a= -\rho v_a  \hspace{10mm} {\rm and} \hspace{10mm} \tilde{\pi}_{ab}= 0\,.  \label{ltrans3}
\end{equation}
Here, $\tilde{\rho}$ is the density, $\tilde{p}$ is the isotropic pressure, $\tilde{q}_a$ is the energy-flux and $\pi_{ab}$ is the anisotropic pressure of the matter, relative to the tilted observers, with $p=0=q_a=\pi_{ab}$ in the CMB frame. Hence, the cosmic medium can be treated as pressure-free in both frames, to first approximation, but not as a perfect fluid. Indeed, in line with (\ref{ltrans3}c), the tilted observers `measure' nonzero effective energy-flux due to their relative motion alone. Also, Eq.~(\ref{ltrans2}) gives
\begin{equation}
\tilde{A}_a= v_a^{\prime}+ {1\over3}\,\Theta v_a\,,  \label{ltrans4}
\end{equation}
since $A_a=0$ in the CMB frame. Therefore, $\tilde{A}_a\neq0$ solely due to the observers peculiar flow.

Essentially all the relative-motion effects discussed in the rest of this paper stem from the fact that $\tilde{q}_a\neq0$ and $\tilde{A}_a\neq0$ (see Eqs.~(\ref{ltrans3}c) and (\ref{ltrans4}c) respectively). The former shows that matter cannot be treated as a perfect fluid in the tilted frame and the latter implies that the worldlines of the tilted observers are no longer geodesics. Recall that $q_a=0=A_a$ in the CMB frame, relative to which the universe is an FRW cosmology filled with a single perfect fluid of zero pressure.\footnote{The fact that $A_a=0$ in the CMB frame and that $\tilde{A}_a\neq0$ in the tilted frame, has immediate and important implications for the form of the associated Raychaudhuri equations. In particular, recalling that $\tilde{\rho}=\rho$ and $\tilde{p}=p=0$ to linear order (see Eqs.~(\ref{ltrans3}a) and (\ref{ltrans3}b) above), we have (in geometrised units)
\begin{equation}
\dot{\Theta}= -{1\over3}\,\Theta^2- {1\over2}\,\rho \hspace{15mm} {\rm and} \hspace{15mm} \tilde{\Theta}^{\prime}= -{1\over3}\,\tilde{\Theta}^2- {1\over2}\,\rho+ \tilde{\rm D}^a\tilde{A}_a\,,  \label{Ray}
\end{equation}
relative to the $u_a$-frame and its tilded counterpart respectively. Given that that $\tilde{A}_a= v_a^{\prime}+(\Theta/3)v_a$ (see Eq.~(\ref{ltrans4})), expression (\ref{Ray}b) shows that relative-motion effects, alone, can change the deceleration/acceleration rate of the tilted observers. This happens in the absence of pressure and  constitutes a significant theoretical difference between our two-fluid study and the `standard' single-fluid approaches. In fact, the aforementioned difference is the reason behind the relative-motion effects discussed in this paper (see \S~\ref{sS-Dq} and \S`\ref{sEq} below).\\ Note that an alternative way of obtaining expression (\ref{pRay1}) is to start from Eqs.~(\ref{Ray}) and then use the linear relations (\ref{dotThetas1}) and (\ref{ltrans4}), while keeping in mind that $\tilde{\vartheta}\ll\Theta$. Finally, it is worth pointing out that combining Eqs.~(\ref{Ray}) with definitions (\ref{qs1}) and then using (\ref{Thetas1}), (\ref{pRay1}) and (\ref{ltrans4}) leads to expression (\ref{qs3}).}
All these mean that linearising around a Friedmann background would generally produce different (linear) equations in the CMB frame and in the tilted frame, due to relative-motion effects alone (e.g.~see footnote~5 and \S~\ref{sRIs} next).

Looking at the first term on the right-hand side of Eq.~(\ref{pRay1}), we notice that the universal expansion always slows the bulk peculiar motion down. In particular, expanding local flows have $\tilde{\vartheta}>0$, in which case the sign of the aforementioned term is negative and therefore decelerates the peculiar expansion. Similarly, when dealing with contracting bulk motions (i.e.~for $\tilde{\vartheta}<0$), the expansion term becomes positive and slows the peculiar contraction down. The role of the second term on the right-hand side of expression (\ref{pRay1}), however, is less straightforward to decode and in order to do so we will turn to cosmological perturbation theory.

\section{The role of the inhomogeneities}\label{sRIs}
Bulk peculiar flows are believed to be the result of the ongoing structure-formation process, which increases the local inhomogeneity and anisotropy of the universe. In fact, one can directly relate the peculiar velocities to inhomogeneities in the density of the matter. In the tilted frame, these are monitored by the dimensionless variable $\tilde{\Delta}_a= (a/\tilde{\rho})\tilde{{\rm D}}_a\tilde{\rho}$, which describes spatial variations in the density between neighbouring (tilted) observers (e.g.~see~\cite{TCM}). Keeping up to $v$-order terms, gives $\tilde{\Delta}_a= (a/\rho)\tilde{{\rm D}}_a\rho$ to linear order. Then, at the same approximation level, this variable evolves as~\cite{TCM}
\begin{equation}
\tilde{\Delta}_a^{\prime}= -\tilde{Z}_a+ {a\Theta\over\rho}\, \left(\tilde{q}_a^{\prime}+{4\over3}\,\Theta\tilde{q}_a\right)- {a\over\rho}\,\tilde{\rm D}_a\tilde{\rm D}^b\tilde{q}_b\,,  \label{tDelta1}
\end{equation}
with $\tilde{Z}_a=a\tilde{\rm D}_a\tilde{\Theta}$ describing inhomogeneities in the volume expansion with respect to the tilted frame.\footnote{In the CMB frame, namely relative to the $u_a$-field, linear inhomogeneities in the density distribution of the (pressureless) matter evolve according to
\begin{equation}
\dot{\Delta}_a= -Z_a\,,  \label{Delta}
\end{equation}
with $\Delta_a=a{\rm D}_a\rho$ and $Z_a=a{\rm D}_a\Theta$ (e.g.~see~\cite{TCM}). Note that, to first approximation, $\Delta_a\neq\tilde{\Delta}_a$ and $Z_a\neq\tilde{Z}_a$}. Recalling that $\tilde{q}_a=-\rho v_a$ in our case (see (\ref{ltrans3}c) above), Eq.~(\ref{tDelta1}) transforms into
\begin{equation}
\tilde{\Delta}_a^{\prime}= -\tilde{Z}_a- a\Theta \left(v_a^{\prime}+{1\over3}\,\Theta v_a\right)+ a\tilde{\rm D}_a\tilde{\vartheta}\,,  \label{tDelta2}
\end{equation}
since $\tilde{\vartheta}=\tilde{\rm D}^av_a$ and $\rho^{\prime}v_a=-\Theta\rho v_a$. Finally, taking the comoving divergence of the above, we arrive at
\begin{equation}
\tilde{\Delta}^{\prime}= -\tilde{Z}- a^2\Theta \left(\tilde{{\rm D}}^av_a^{\prime} +{1\over3}\,\Theta\tilde{\vartheta}\right)+ a^2\tilde{\rm D}^2\tilde{\vartheta}\,,  \label{tDelta3}
\end{equation}
where $\tilde{\Delta}=a\tilde{\rm D}^a\tilde{\Delta}_a$ and $\tilde{Z}=a\tilde{\rm D}^a\tilde{Z}_a$. Also, $a\tilde{\rm D}^a\tilde{\Delta}_a^{\prime}=\tilde{\Delta}^{\prime}$ to linear order. By construction, the scalar $\tilde{\Delta}$ monitors overdensities or underdensities in the matter distribution, relative to the tilted frame (e.g.~see~\cite{TCM}). Solving Eq.~(\ref{tDelta3}) for $\tilde{\rm D}^av_a^{\prime}$ and substituting the resulting expression into the right-hand side of (\ref{pRay1}) gives
\begin{equation}
\tilde{\vartheta}^{\prime}= -{2\over3}\,\Theta\tilde{\vartheta}+ {1\over\Theta}\, \tilde{\rm D}^2\tilde{\vartheta}- {1\over a^2} \left({\tilde{\Delta}^{\prime}\over\Theta} +{\tilde{Z}\over\Theta}\right)\,.  \label{pRay2}
\end{equation}
This result provides the linear evolution formula of $\tilde{\vartheta}$, namely the linearised peculiar Raychaudhuri equation (see~\cite{TK} for a more general expression), written in the frame of the tilted observers. Note the Laplacian term on the right-hand side of Eq.~(\ref{pRay2}), which introduces a scale-dependence that will prove crucial later.

Our next step is to harmonically decompose the perturbations, by setting $\tilde{\vartheta}= \Sigma_n\tilde{\vartheta}_{(n)}\mathcal{Q}^{(n)}$, $\tilde{\Delta}= \Sigma_n\tilde{\Delta}_{(n)}\mathcal{Q}^{(n)}$ and $\tilde{Z}= \Sigma_n\tilde{Z}_{(n)}\mathcal{Q}^{(n)}$, where $\tilde{\rm D}_a\tilde{\vartheta}_{(n)}=0=\tilde{\rm D}_a \tilde{\Delta}_{(n)}=\tilde{\rm D}_a\tilde{Z}_{(n)}$ and $n$ represents the comoving eigenvalue of the harmonic mode. Also, $\mathcal{Q}^{(n)}$ are standard scalar harmonic functions, with $\mathcal{Q}^{\prime(n)}=0$ and $\tilde{\rm D}^2\mathcal{Q}^{(n)}=-(n/a)^2\mathcal{Q}^{(n)}$. On using all of the above, expression (\ref{pRay2}) takes the form
\begin{equation}
\tilde{\vartheta}_{(n)}^{\prime}= -{2\over3}\,\Theta\tilde{\vartheta}_{(n)}- \left({n\over a}\right)^2{\tilde{\vartheta}_{(n)}\over\Theta}- {1\over a^2} \left({\tilde{\Delta}_{(n)}^{\prime}\over\Theta} +{\tilde{Z}_{(n)}\over\Theta}\right)\,.  \label{pRay3}
\end{equation}
Finally, since $\lambda_n=a/n$ is the physical scale of the perturbation (i.e.~the size of the bulk flow), $\lambda_H=1/H= 3/\Theta$ is the Hubble radius and $\lambda_K=a/|K|$ (with $K=\pm1$) is the curvature scale of the universe, we arrive at the expression
\begin{equation}
{\tilde{\vartheta}_{(n)}^{\prime}\over\Theta^2}= -{2\over3} \left[1+{1\over6}\left({\lambda_H\over\lambda_n}\right)^2\right] {\tilde{\vartheta}_{(n)}\over\Theta}- {1\over9}\left({\lambda_H\over\lambda_K}\right)^2 \left({\tilde{\Delta}_{(n)}^{\prime}\over\Theta} +{\tilde{Z}_{(n)}\over\Theta}\right)\,.  \label{pRay4}
\end{equation}
which shows that the value of the dimensionless ratio $\tilde{\vartheta}_{(n)}^{\prime}/\Theta^2$ is scale dependent. In fact, there are three scale-related thresholds in Eq.~(\ref{pRay4}), corresponding to the size of the bulk flow ($\lambda_n$), to the Hubble radius ($\lambda_H$) and to the curvature scale ($\lambda_K$) of the universe. What is perhaps more important is that, due to the scale dependence of the above relation, small values of $\tilde{\vartheta}_{(n)}/\Theta$ do not necessarily imply the same for the ratio $\tilde{\vartheta}_{(n)}^{\prime}/\Theta^2$, as seen from the tilted frame. When dealing with bulk flows much smaller than the horizon (i.e.~those with $\lambda_H/\lambda_n\gg1$), in particular, the smallness of $\tilde{\vartheta}_{(n)}/\Theta$ does not guarantee the same for $\tilde{\vartheta}_{(n)}^{\prime}/\Theta^2$.

\section{The scale-dependence of $\tilde{q}$}\label{sS-Dq}
By means of Eq.~(\ref{qs3}), the scale dependence of $\tilde{\vartheta}_{(n)}^{\prime}$ passes on to the deceleration parameter, as measured in the tilted frame. More specifically, substituting (\ref{pRay4}) into the right-hand side of (\ref{qs3}), gives
\begin{equation}
1+ \tilde{q}_{(n)}= (1+q)\left(1-2{\tilde{\vartheta}_{(n)}\over\Theta}\right)+ 2\left[1+{1\over6}\left({\lambda_H\over\lambda_n}\right)^2\right] {\tilde{\vartheta}_{(n)}\over\Theta}+ {1\over3}\left({\lambda_H\over\lambda_K}\right)^2 \left({\tilde{\Delta}_{(n)}^{\prime}\over\Theta} +{\tilde{Z}_{(n)}\over\Theta}\right)\,,  \label{qs4}
\end{equation}
to first approximation. Before proceeding to extract some representative numerical estimates for $\tilde{q}_{(n)}$, let us first look at the last term on the right-hand side of the above. Today, observations strongly indicate that $(\lambda_H/\lambda_K)^2=|1-\Omega_K|\lesssim10^{-2}$. In addition, $\tilde{Z}_{(n)}/\Theta\ll1$ due to the anticipated large-scale homogeneity of the universe and the same could also be argued for the dimensionless ratio $\tilde{\Delta}_{(n)}^{\prime}/\Theta$. Then, expression (\ref{qs4}) reduces to\footnote{Strictly speaking, the observations restrict the density contrast $\tilde{\Delta}_{(n)}$ well bellow unity, but not its temporal derivative and therefore the dimensionless ratio $\tilde{\Delta}_{(n)}^{\prime}/\Theta$. Nevertheless, given that $(\lambda_H/\lambda_K)^2 \lesssim10^{-2}$, the product $(\lambda_H/\lambda_K)^2 (\tilde{\Delta}_{(n)}^{\prime}/\Theta)$ should be negligible on all scales of interest. Put another way, $(\lambda_H/\lambda_K)^2 (\tilde{\Delta}_{(n)}^{\prime}/\Theta)\ll1$, unless $\tilde{\Delta}_{(n)}^{\prime}/\Theta\gtrsim10^2$ in regions of few hundred Mpc (or larger). The latter seems rather unlikely.}
\begin{equation}
1+ \tilde{q}_{(n)}\simeq (1+q)\left(1-2{\tilde{\vartheta}_{(n)}\over\Theta}\right)+ 2\left[1+{1\over6}\left({\lambda_H\over\lambda_n}\right)^2\right] {\tilde{\vartheta}_{(n)}\over\Theta}\,,  \label{qs5}
\end{equation}

Observations of bulk peculiar motions typically cover scales of few hundred Mpc~\cite{WFH,CMSS}, although there have been reports of peculiar flows on Gpc~scales~\cite{KA-BKE}. In all cases, the size of the drifting region lies inside the horizon (i.e.~$\lambda_n<\lambda_H$) and the value of the peculiar velocity is a small fraction of the Hubble speed on the corresponding scale (i.e.~$\tilde{\vartheta}_{(n)}/ \Theta\ll1$). On these wavelengths, expression (\ref{qs5}) simplifies to
\begin{equation}
\tilde{q}_{(n)}\simeq q+ {1\over3}\left({\lambda_H\over\lambda_n}\right)^2 {\tilde{\vartheta}_{(n)}\over\Theta}\,,  \label{qs6}
\end{equation}
providing a very simple relation between the deceleration parameter of the actual universe ($q$) and the one measured in the rest-frame of a typical galaxy ($\tilde{q}_{(n)}$), which resides within a bulk flow of size $\lambda_n$ that moves relative to the smooth Hubble expansion.

As expected, in the CMB frame where all peculiar velocities vanish identically, Eq.~(\ref{qs5}) confirms that $\tilde{q}=q$ on all scales, as expected. In the rest frame of the bulk motion, however, $\tilde{q}\neq q$ and the difference also depends on the scale of the bulk flow. It comes to no surprise that the effect of the peculiar motion drops with increasing scale. In fact, on superhorizon lengths (with $\lambda_H/\lambda_n\ll1$) we find that $\tilde{q}\rightarrow q$, which is something to be expected as well. The scale dependence seen in Eq.~(\ref{qs6}) becomes the decisive factor as we move inside the Hubble radius and towards successively smaller lengths. In fact, on sufficiently small scales (i.e.~for $\lambda_H/ \lambda_n\gg1$), even relatively slow bulk motions (i.e.~those with $\tilde{\vartheta}_{(n)}/ \Theta\ll1$) can have a measurable effect on the deceleration parameter. Note, however, that one should not apply expression (\ref{qs5}) to scales much smaller than 100~Mpc, since there the peculiar velocities are a considerable fraction of the Hubble speed (i.e.~$\tilde{\vartheta}_{(n)}/\Theta\sim1$) and our approximation breaks down.

The most intriguing feature of Eq.~(\ref{qs6}) is that the relative motion might even change the sign of the deceleration parameter. This could happen when the peculiar flow is contracting (i.e.~for $\tilde{\vartheta}_{(n)}<0$) and from now on, we will focus on contracting bulk flows only. Following our original assumption, that relative to the CMB frame the universe is an FRW model, we may set $q=1/2$ in the right-hand side of Eq.~(\ref{qs6}). Also, since the background geometry is essentially Euclidean and the peculiar velocities are non-relativistic, we may write $|\tilde{\vartheta}|\simeq\partial^av_a\simeq3\langle v\rangle/\lambda$. Here, $\langle v\rangle$ is the mean magnitude of the peculiar velocity and $\lambda=\lambda_n$ is the size of the associated bulk motion (we have dropped the $n$-index for the economy of the presentation). Then, given that $\Theta=3H$, Eq.~(\ref{qs6}) can be written as
\begin{equation}
\tilde{q}_{_{\lambda}}\simeq {1\over2}- {1\over3}\left({\lambda_H\over\lambda}\right)^2 {\langle v\rangle\over\lambda H}\,,  \label{qs7}
\end{equation}
since we have assumed contracting peculiar flows (with $\tilde{\vartheta}/\Theta\simeq-\langle v\rangle/\lambda H$). Note also that $\tilde{q}_{\lambda}=\tilde{q}_{(n)}$ in our new notation. Substituting into the right-hand side of the above the size and the mean velocity of a bulk flow, provides the value of the deceleration parameter measured by observers living within the flow.

\section{Estimating $\tilde{q}$}\label{sEq}
In some cases the average peculiar velocity of the bulk flow may not be available. Also, sometimes the divergence of the observed mean bulk velocity may not accurately provide the volume scalar of the flow. It is conceivable, in particular, that $|\tilde{\vartheta}|\lesssim3\langle v\rangle/\lambda$. One could account for this by introducing a parameter ($\alpha$), so that $\tilde{\vartheta}=-3\alpha\langle v\rangle/\lambda$ for contracting bulk flows, and then recasting expression (\ref{qs7}) into
\begin{equation}
\tilde{q}_{_{\lambda}}\simeq {1\over2}- {1\over3}\left({\lambda_H\over\lambda}\right)^2 {\alpha\langle v\rangle\over\lambda H}\,,  \label{qs8}
\end{equation}
with $0<\alpha<1$ in most cases. Next, we will employ the above and use peculiar-velocity measurements from recent surveys of large-scale bulk motions to obtain numerical estimates for the values of the deceleration parameter measured inside these flows.

Following (\ref{qs8}), the value of $\tilde{q}_{\lambda}$ depends stronger on the scale of the bulk motion than on its speed. In particular, the effects of relative motion on $\tilde{q}_{\lambda}$ are more pronounced on scales well inside the Hubble radius. Such large-scale peculiar flows have been recently reported by several groups. Most of the surveys cover lengths close to $100/h\simeq150$~Mpc, though they have been claims of coherent flows with radii of Mpc-order (the so-called dark flows). The latter reports seem to be at odds with the data collected from the Planck mission (see however~\cite{A-B}), but peculiar motions on scales smaller than 400~Mpc are not constrained by Planck. These surveys seem to agree on the direction of the bulk flow, but vary on the magnitude of the peculiar velocity. Some of the reported magnitudes are close to the general expectations of the $\Lambda$CDM scenario, though others are in excess of the predicted values. Here, we use representative peculiar-velocity magnitudes of large-scale bulk motions, to estimate the value of the deceleration parameter measured by observers living within the flow. In all cases we will assume that $q=1/2$, $H\simeq70$~km/sec\,Mpc and $\lambda_H\simeq3\times10^3$~Mpc.

The most `conservative' claims are probably those by Nusser and Davis, reporting mean peculiar velocities close to $\langle v\rangle\simeq 250$~km/sec on scales of roughly $\lambda\simeq150$~Mpc~\cite{ND}. This data, which are in agreement with the $\Lambda$CDM constraints, correspond to $(\lambda_H/\lambda)\simeq20$ and $\langle v\rangle/\lambda H\simeq0.02$. Substituting these values into the right-hand side of Eq.~(\ref{qs8}) and assuming that $\alpha=1$, namely that $|\tilde{\vartheta}|=3\langle v\rangle/ \lambda$, we find that $\tilde{q}\simeq-2.16$ on the associated scale (see Table~1). When $|\tilde{\vartheta}|<3\langle v\rangle/ \lambda$, the value of the local deceleration parameter increases. Setting $\alpha=2/3$ and $\alpha=1/2$, for example, gives $\tilde{q}=-1.27$ and $\tilde{q}=-0.83$ respectively. On the same scales, Lavaux~et~al and Macaulay~et~al have reported larger peculiar velocities (around 370~km/sec)~\cite{LTMC}. Using this survey, the deceleration parameter can go as low as $\tilde{q}\simeq-3.50$. Much lower values for $\tilde{q}$ are obtained from the surveys of Watkins~et~al~\cite{WFH}, reporting peculiar velocities close to 400~km/sec on scales of approximately 150~Mpc (see Table~1).

We should emphasise again that none of the aforementioned values of the deceleration parameter are `real', but they are simply the result of the observers' relative motion with respect to the smooth Hubble expansion. In the CMB frame, the peculiar velocities vanish identically and $\tilde{q}=q=1/2$, as expected in an FRW universe. Intuitively speaking, observers residing inside a contracting bulk flow could be misled to believe that the background expansion accelerates, in the same way passengers in a vehicle that slows down might think that the overtaking cars are speeding up.

\begin{table}
\caption{Representative values of the deceleration parameter ($\tilde{q}$), based on the peculiar velocities ($\langle v\rangle$) reported from several bulk-motion surveys, as measured by observers living inside these flows. Note that the `background' universe is a decelerating Friedmann model (with $q=1/2$) and in the observers' rest-frame the value of the deceleration parameter changes entirely due to relative-motion effects. The numerical estimates follow from Eq.~(\ref{qs8}), for two different values of the $\alpha$-parameter, assuming that $H\simeq70$~km/sec\,Mpc and $\lambda_H\simeq3\times10^3$~Mpc. Note that the data from Nusser \& Davis (first line) are in agreement with the $\Lambda$CDM model.}
\vspace{0.5truecm}
\begin{center}\begin{tabular}{cccccccc}
\hline \hline & Survey & $\lambda~({\rm Mpc})$ & $\langle v\rangle~({\rm km/s})$ & $\alpha=1$ & $\alpha=1/2$ &\\ \hline \hline & $\begin{array}{c} {\rm Nusser\,\&\,Davis\;(2011)} \\ {\rm Lavaux,\,et\,al\;(2010)} \\ {\rm Watkins,\,et\,al\;(2009)} \\ {\rm Colin,\,et\,al\;(2011)} \\ {\rm Abate\,\&\,Feldman\;(2012)} \\ {\rm Ade,\,et\,al\;(2014)} \end{array}$ & $\begin{array}{c} 150 \\ 150 \\ 150 \\ 250 \\ 1000 \\ \gtrsim500 \end{array}$ & $\begin{array}{c} 250 \\ 370 \\ 400 \\ 260 \\ 4000 \\ \lesssim400 \end{array}$ & $\begin{array}{c} \tilde{q}_{_{\lambda}}\simeq-2.16 \\ \tilde{q}_{_{\lambda}}\simeq-3.50 \\ \tilde{q}_{_{\lambda}}\simeq-4.83 \\
\tilde{q}_{_{\lambda}}\simeq-0.22 \\
\tilde{q}_{_{\lambda}}\simeq+0.32 \\
\tilde{q}_{_{\lambda}}\gtrsim+0.38 \end{array}$ & $\begin{array}{c} \tilde{q}_{_{\lambda}}\simeq-0.83 \\ \tilde{q}_{_{\lambda}}\simeq-1.50 \\ \tilde{q}_{_{\lambda}}\simeq-2.16 \\
\tilde{q}_{_{\lambda}}\simeq+0.14 \\
\tilde{q}_{_{\lambda}}\simeq+0.41 \\ \\ \end{array}$\\ [2.5truemm] \hline \hline
\end{tabular}\end{center}\label{tab1}\vspace{0.5truecm}
\end{table}

Following (\ref{qs8}), the impact of the relative motion on $\tilde{q}$ weakens significantly as we move on to progressively larger wavelengths, where the ratio $(\lambda_H/\lambda)^2$ drops sharply. Colin~et~al, for example have reported peculiar velocities around 260~km/sec on scales close to 250~Mpc~\cite{CMSS}. Substituting the findings of this survey into the right-hand side of Eq.~(\ref{qs8}), we find that $\tilde{q}$ can take only marginally negative values. Also, on Gpc-order scales, Kashlinsky~et~al and Abate~et~al have reported bulk motions as fast as 1000~km/sec and 4000~km/sec respectively~\cite{KA-BKE}. Nevertheless, despite the large amplitude of the peculiar velocities, their effect on the value of the deceleration parameter, as measured within these flows, is relatively weak. Using the velocities reported by Abate~et~al, in particular, expression (\ref{qs8}) gives $\tilde{q}\simeq +0.32$ on the corresponding lengths (for $\alpha=1$ -- see Table~1).

Finally, the analysis of the Planck data seems to set an upper limit of $\langle v\rangle\simeq400$~km/sec on the mean velocity of bulk motions larger than $350/h\simeq500$~Mpc~\cite{Aetal}. Using this constraint, Eq.~(\ref{qs8}) puts a lower limit of $\tilde{q}\simeq+0.38$ on the corresponding scale (see last line in Table~1).

\section{Discussion}\label{sD}
Observers moving relative to each other could have a different view of `reality', and they might also interpret their observations differently. In this study we have assumed two families of observers with a non-relativistic peculiar velocity between them. The first family was identified with the smooth expansion of a dust-dominated Friedmann universe and the second (the so-called `tilted' observers) with a typical galaxy moving relative to the Hubble flow. Our aim was to investigate whether relative motion modifies the kinematics of the drifting observers. More specifically, we wanted to look for differences in the value of the deceleration parameter, as measured in the aforementioned two frames, that were solely due to relative-motion effects. We were primarily motivated by resent surveys, reporting coherent bulk motions with sizes of few hundred Mpc and peculiar velocities in excess of those anticipated by the $\Lambda$CDM paradigm. In agreement with~\cite{T1,T2}, the deceleration parameter in the rest-frame of large-scale bulk flows was generally found to differ from that of the background universe. Extending that work, we  realised that the aforementioned difference also depends on the scale of the bulk motion and not only on the magnitude of the peculiar velocity. In particular, the smaller the size of the drifting domain (relative to the Hubble radius), the stronger the relative-motion effects on the local deceleration parameter. Moreover, observers inside slightly contracting bulk flows could measure a negative deceleration parameter in their own rest-frame, while the universe they live in is actually decelerating.

None of the aforementioned  effects is real of course, given that they all vanish in the CMB frame of the smooth Hubble expansion. Just like the CMB dipole, all of the above are nothing else but mere artifacts of the observers' relative motion. Nevertheless, when not properly taken into account, such relative-motion effects could potentially lead to the misinterpretation of the observations. To avoid this, one needs to abandon the single-fluid models and allow for a second set of observers moving relative to the Hubble flow. It is then straightforward to show that to these observers the cosmic medium no longer looks like a perfect fluid, even when their peculiar velocity is small. As a result, the Raychaudhuri equation in the rest-frame of a large-scale bulk motion has an additional term, which carries the relative-motion effects (see Eq.~(\ref{Ray}b) in footnote~4). The latter are immediately transferred to the local deceleration parameter (see Eqs.~(\ref{qs4})-(\ref{qs8})), making it different from that of the actual universe (sometimes even in the sign).

Following Table~1, the deceleration parameter measured inside slightly contracting bulk motions of few hundred Mpc is particularly voulnerable to relative motion effects. Adopting the peculiar velocities claimed by Watkins et al, which are in tension with the $\Lambda$CDM paradigm, we obtained values as low as $\tilde{q}\simeq-4.83$ on scales around 150~Mpc. On the same wavelengths, the more conservative bulk velocities reported by Nusser et al, which are within the $\Lambda$CDM limits, gave a lowest value of $\tilde{q}\simeq-2.16$. Analogous numerical estimates for the deceleration parameter were claimed in~\cite{GW}. There, the authors used a phenomenological two-parameter function for their deceleration parameter to reconstruct its evolution over the last couple of billion years, assuming a dark-energy dominated (exact) FRW universe,. Here, we have arrived at a qualitatively similar result, working in a tilted (perturbed) Friedmannian cosmology with pressure-free matter. Finally, according to (\ref{qs8}), the effects of relative motion on the deceleration parameter inside the bulk flow, drop sharply with increasing scale. This means that $\tilde{q}$ quickly approaches its `background' value (i.e.~$\tilde{q}\rightarrow q=1/2$), as we move on to progressively longer wavelengths. To the unsuspecting observer living inside the bulk flow, however, this may give the false impression that the universal acceleration is a relatively recent event.

Is it then possible that the recent accelerated expansion of the universe could be a mere illusion and a relative-motion artifact? Although this may not be inconceivable, in principle, it is also rather premature to make such a claim. What we could say at this point is that there are many attractive aspects in this alternative, assuming that it works. There is no need for dark energy, it is not necessary to modify gravity and there is no reason to abandon the Friedmann models. Everything happens within standard physics and conventional cosmology. In addition, since on the scales of interest the observed peculiar velocities are much smaller than the Hubble speed, we have not yet included any nonlinear effects. There seems to be a key requirement however. In order to create the (false) impression of accelerated expansion, the bulk flow we live in must be slightly contracting (on a background that is expanding). In a sense, observers living inside a contracting bulk flow may think that the background expansion is accelerating, just like passengers in a car that slows down are sometimes mislead to believe that the overtaking vehicles are speeding up. At the moment, we cannot measure volume changes (i.e.~the value and the sign of $\tilde{\vartheta}$) for bulk motions of 100~Mpc-order or greater. Nevertheless, provided there is no bias in favour of contraction or expansion on these wavelengths, the chances of living within a slightly contracting large-scale peculiar flow should be around 50\,\%.

There are caveats as well of course. For example, we have assumed that the volume scalar ($\tilde{\vartheta}$) of the bulk flow can be obtained from the its observed mean velocity ($\langle v\rangle$), by means of the approximate formula $\tilde{\vartheta}\simeq3\alpha\langle v\rangle/\lambda$, with $0<\alpha<1$. It is conceivable, however, that the $\alpha$-parameter might take very small values. For instance, if the bulk flow is incompressible, then $\alpha,\,\tilde{\vartheta}\rightarrow0$ and the relative-motion effects on $\tilde{q}$ become negligible for all practical purposes. A more subtle caveat lies perhaps in our assumption that the ratios $\tilde{\Delta}_{(n)}^{\prime}/\Theta$ and $\tilde{Z}_{(n)}/\Theta$ inside the last term of Eq.~(\ref{qs4}) are much smaller than unity and therefore their effect on $\tilde{q}$ is irrelevant. It is not a priori inconceivable, however, that in the rest-frame of the bulk motion these ratios may not be small and that their effect on $\tilde{q}$ may not be negligible. In particular, the density contrast $\tilde{\Delta}$ could vary rapidly with time in the tilted frame, in which case the ratio $\tilde{\Delta}^{\prime}/\Theta$ may take relatively large values. Having said that, the coefficient $(\lambda_H/\lambda_K)^2$ of the aforementioned last term is very small. More specifically, $(\lambda_H/\lambda_K)^2\lesssim10^{-2}$, according to the current observational constrains. Therefore, unless $\tilde{\Delta}_{(n)}^{\prime}/\Theta+ \tilde{Z}_{(n)}/\Theta\gtrsim10^2$ in the tilted frame, the last term on the right-hand side of Eq.~(\ref{qs4}) will be irrelevant. It should also be noted that, if the inhomogeneous (i.e.~the last) term dominates the right-hand side of (\ref{qs4}), the value of the deceleration parameter in the tilted frame will still differ from that of the actual universe (i.e.~$\tilde{q}\neq q$). In that case, however, the conditions leading to negative $\tilde{q}$ may not be the same with the ones discussed in \S~\ref{sS-Dq} and \S~\ref{sEq} earlier. Overall, the only way of having  $\tilde{q}\simeq q$, is when the last two terms on the right-hand side of Eq.~(\ref{qs4}) essentially cancel each other out. It is conceivable that some of the these issues may be resolved when nonlinear effects are included into the study.

We would like to close our discussion by pointing out that we have assumed observers with individual peculiar velocities equal to the average velocity of the bulk flow. In general. however, a typical observer inside the bulk will have peculiar velocity close (in terms of both magnitude and direction) to the mean, but not equal to it. Put another way, typical observers will have their own (small) peculiar velocities relative to the average motion of the bulk flow. To these observers, the distribution of $\tilde{q}$ will not be entirely isotropic, but it should demonstrate a weak dipole-like anisotropy analogous to that seen in the CMB. Recall that the latter is caused by the motion of the bulk with respect to the smooth universal expansion. In a similar way, the aforementioned $\tilde{q}$-anisotropy will be due to the observer's peculiar flow relative to the mean motion of the bulk. Assuming that the difference between the two  velocities is small, the anisotropy in the observed distribution of $\tilde{q}$ will be weak and the associated dipole axis should lie close to that of the CMB (see~\cite{T2} for further discussion). Interestingly, there have been surveys arguing that such a weak dipolar anisotropy might actually exist in the supernovae data. In other words, there have been claims suggesting that our universe is accelerating slightly faster in one direction and equally slower along the opposite (e.g.~see~\cite{CL-B}). Moreover, the dipole axis seems to lie fairly close to that of the CMB. If these reports are confirmed, one might want to take a closer look at the scenario outlined here as an alternative explanation of the recent universal acceleration. So far, however, the presence of a weak dipolar axis in the supernovae data remains ambiguous.

\end{document}